\documentclass[10pt,conference,compsocconf,letterpaper]{IEEEtran}
\IEEEoverridecommandlockouts
\usepackage{cite}
\usepackage{balance}
\usepackage{mathrsfs}
\usepackage{amsmath,amssymb,amsfonts,amsthm}
\usepackage{enumitem}
\usepackage{amsmath}
\usepackage{graphicx}
\usepackage{extarrows}
\usepackage{amssymb}
\usepackage{subfigure}
\usepackage{verbatim}
\usepackage{makecell}
\usepackage{diagbox}
\usepackage{bm}
\usepackage{multirow}
\usepackage{float}
\usepackage{placeins}
\usepackage[linesnumbered, ruled]{algorithm2e}
\SetAlFnt{\small}
\usepackage{epstopdf}
\usepackage{setspace}
\usepackage[super]{nth}
\usepackage{fancyhdr}
\usepackage{slashbox}
\usepackage{lipsum,multicol}
\usepackage{url}
\usepackage{xcolor}
\usepackage{booktabs}
\usepackage{diagbox}
\usepackage{tikz}
\usepackage{graphicx}
\usepackage{graphbox}
\usepackage{adjustbox}
\usepackage[T1]{fontenc}
\usepackage[font=footnotesize, skip=3pt]{caption}
\usepackage{pifont}

\allowdisplaybreaks[4]

\def \eg {\emph{e.g.}, }
\def \ie {\emph{i.e.}, }

\usepackage[linesnumbered, ruled]{algorithm2e}
\def\BibTeX{{\rm B\kern-.05em{\sc i\kern-.025em b}\kern-.08em
    T\kern-.1667em\lower.7ex\hbox{E}\kern-.125emX}}

\SetCommentSty{mycommfont} 

\newcommand{\model}{{Nova}\xspace}

\newcommand{\highlight}[1]{#1}

\newcommand{\blackcircletext}[1]{%
  \tikz[baseline=(char.base)]{
    \node[shape=circle, draw=black, fill=black, text=white, inner sep=1pt] (char) {\textbf{#1}};
  }%
}

\usepackage[firstpage]{draftwatermark}
\SetWatermarkText{\hspace*{7.56in}\raisebox{10.1in}
  {\includegraphics[scale=0.1]{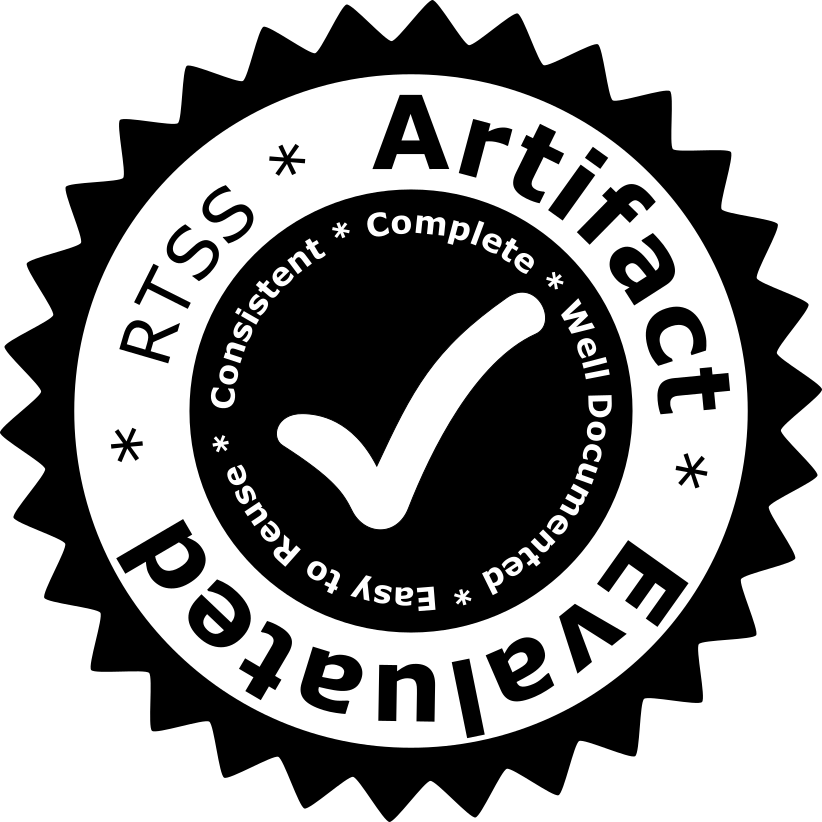}}}
\SetWatermarkAngle{0}
\usepackage{url}

\begin{document}
\title{Nova: Real-Time Agentic Vision-Language Model Serving with Adaptive Cross-Stage Parallelization}

\author{Yuhang Xu$^{\dagger}$, Shengzhong Liu$^{*\dagger}$, Dong Zhang$^{\S}$, Bingheng Yan$^{\S}$, Fan Wu$^{\dagger}$, Guihai Chen$^{\dagger}$ \thanks{*Shengzhong Liu is the corresponding author.}\\
$^{\dagger}$Shanghai Jiao Tong University
, $^{\S}$Inspur Data Co.,Ltd.\\
Email: \{xuyuhangtmx, shengzhong\}@sjtu.edu.cn,  
\{zhangdong, yanbh\}@inspur.com, \\
\{fwu, gchen\}@cs.sjtu.edu.cn}

\maketitle

\begin{abstract}
This paper presents \model, a real-time scheduling framework for serving agentic vision-language models (VLMs) on a single GPU with balanced per-request latency and overall request process throughput.
Our design begins by enabling effective pipelining across vision encode, LLM prefill, and LLM decode stages of VLMs, by exploiting their heterogeneous resource demands during execution and incorporating elastic GPU spatial partitioning among stages to maximally utilize the compute and memory resources. 
Building on this, we introduce a real-time scheduling algorithm that adaptively calibrates resource allocation among stages based on a Pareto-optimal analysis of the latency–throughput trade-off, allowing the system to sustain responsiveness and resource efficiency under dynamic request loads. 
To further alleviate GPU memory pressure, we design a lightweight weight offloading strategy for vision encoders that preserves inference efficiency with minimized memory overhead. 
Extensive evaluations on both synthetic and real-world agent workloads demonstrate that \model consistently outperforms the state-of-the-art baselines, improving the maximum latency by up to 23.3\%, while keeping competitive throughput.
\end{abstract}

\section{Introduction}\label{sec:intro}

Recent advances in vision-language models (VLMs) have enabled a new class of interactive agents that can perceive visual environments and generate language-based actions~\cite{wang2024mobile_agent, hong2024cogagent, zhang2024vision_vlm_survey, zhu2023minigpt, niu2024screenagent, zhai2024fine_vlm_decision_making, zhao2025responsive_mocha}. Such VLM agents, capable of interpreting graphical user interfaces (GUIs) and textual user instructions, offer transformative capabilities across application domains, including mobile automation, remote control, and digital assistance~\cite{li2024personal_llm_agent, ferrag2025llm_autonomos_ai_agent, dong2024survey_llm_based_agent}. Different from standard LLM applications like chat and summarization, VLM agents must operate under tight latency constraints to ensure smooth and reactive behaviors. As a result, responsive serving becomes a critical requirement for deploying VLM-based agents in practice.

\highlight{This paper considers the scenario of serving agentic VLMs on a single GPU, designed for data-sensitive application domains (\eg banking, healthcare, government) that prohibit offloading GUI images to cloud servers.} Unlike cloud clusters consisting of massive GPUs, where different stages of VLM inference, including vision encode, LLM prefill, and auto-regressive LLM decode, may be distributed across separate GPUs~\cite{zhong2024distserve, patel2024splitwise}, edge serving executes all stages on a single device. These stages exhibit heterogeneous resource demands, inducing significant challenges to GPU resource management. Besides, the short but bursty nature of agentic workloads makes it difficult to balance per-request latency with overall system throughput.

Heterogeneous VLM stages pose significant challenges to balancing throughput and latency, as prioritizing the vision encode or LLM prefill can severely slow down token generation, while favoring LLM decode may lead to servere GPU under-utilization. 
To address this issue, techniques like chunked prefill~\cite{agrawal2024taming_sarathi_chunkprefill} are proposed to divide LLM prefill into multiple encode iterations by encoding one chunk at a time, parallelize prefill with decode in a hybrid batch, and leverage their shared model architecture and weights with homogeneous operators, such that later requests in the waiting queue can receive their first response within shorter time to first token (TTFT). 
However, they fall short in supporting agentic VLM workloads, as the vision encoder and LLM modules are structurally separated, hindering data-level cross-stage batching. Furthermore, vision encode takes significantly longer than LLM prefill, \eg over 2$\times$ in our measurements, making it the new bottleneck beyond batching. 
Even operator-level optimizations like POD-Attention~\cite{kamath2025pod-attention} that do not require shared weights become ineffective for overlapping vision and language stages. 
Under high request load, this mismatch can cause even decode-prioritized approaches, such as chunked prefill, to suffer from decode starvation and resource underutilization.

These observations motivate us to rethink the system design for agentic VLM serving. To this end, we propose \model, a pipeline-parallel, resource-aware serving framework designed to address the stage heterogeneity and latency sensitivity of agentic VLM inference. We begin by estimating the feasibility of cross-stage pipelining through multi-kernel co-execution. Unlike operator-level batching depending on shared model weights~\cite{agrawal2024taming_sarathi_chunkprefill, kamath2025pod-attention}, kernel co-execution allows concurrent execution across structurally independent stages. While prior work, such as NanoFlow~\cite{zhu2024nanoflow}, enables such parallelism via fine-grained kernel re-implementation and customized scheduling, we adopt a hardware-centric approach by exploring GPU spatial sharing~\cite{yu2020fine_grained_gpu_sharing, strati2024orion, wu2023transparent_tgs} as a more generalizable and elastic solution.
Our key insight is that the primary barrier to parallelization arises from resource contention between heterogeneous kernel launches. To address this, we exploit streaming multiprocessor (SM) partitioning to enable multi-stage co-execution on the GPU. 
Specifically, vision encode and LLM decode often contend for limited registers and shared memory, which inhibits concurrent kernel launches despite their complementary compute and memory demands. By assigning disjoint SM subsets to these inference stages, their contention is alleviated and co-execution efficiency is significantly improved.

Previous GPU sharing efforts have primarily focused on maximizing the overall system throughput~\cite{wu2023transparent_tgs, zhao2021exploiting_intra_sm} but overlooked per-request inference latency. 
We instead propose an adaptive SM allocation strategy that navigates the trade-off between these two metrics in real time. 
Unlike static partitioning that fails to accommodate real-time workload variations, we model the resource demands of each stage and dynamically calibrate GPU allocation based on the runtime request load. 
Specifically, we leverage Pareto-optimal analysis to guide SM partitioning that minimizes end-to-end latency across varying conditions. This is particularly crucial under bursty request patterns, where decode batching and front-stage congestion must be jointly managed. Therefore, SM partitioning not only achieves pipeline parallelism to reduce latency but also exposes resource-level concurrency to improve throughput. 
The stage-level scheduling aligns better with the modular VLM structures than operator-level techniques and achieves consistent improvement across varying workloads.

We further identify the GPU memory optimization space in vision encoder inference in agentic VLMs, which compete with the pre-allocated KV cache of LLM decode.
This contention can lead to cache eviction or recomputation~\cite{kwon2023efficient_page_attn}, which is problematic in memory-constrained serving scenarios. To address this, we introduce a lightweight weight offloading mechanism that asynchronously swaps vision encoder layers between CPU and GPU memory. This approach alleviates memory pressure with negligible latency overhead, preserving KV cache space and improving system generalizability across hardware configurations.

We implement \model and evaluate its performance on multiple platforms using both synthetic workloads and real-world workloads. The results show that \model consistently achieves lower end-to-end latency, outperforming the SOTA baselines by up to 14.6\% in average latency and 23.3\% in maximum latency, while sustaining no smaller throughput.  

Overall, our main contributions are summarized as:
\begin{itemize}
    \item We identify and address the unique system challenges in serving agentic VLMs on a single GPU, which involve heterogeneous and stage-wise workloads different from standard LLM serving systems;
    \item We propose a stage-parallel execution framework based on inter-SM GPU sharing, combined with an adaptive SM partitioning strategy guided by Pareto-optimal analysis, to dynamically balance per-request latency and overall throughput under varying workloads;
    \item We design a lightweight weight offloading mechanism for large vision encoders, mitigate GPU memory pressure with minimal overhead;
    \item We conduct extensive experiments using both synthetic and real-world datasets on multiple platforms to demonstrate the effectiveness and scalability of our design.
\end{itemize}

\section{Background and Motivations}\label{sec:background_motivation}

\subsection{GUI-based Agentic Vision-Language Model}
\label{sec:vlm_agent}
\begin{figure}[t!]
    \centering
    \includegraphics[width=0.95\linewidth]{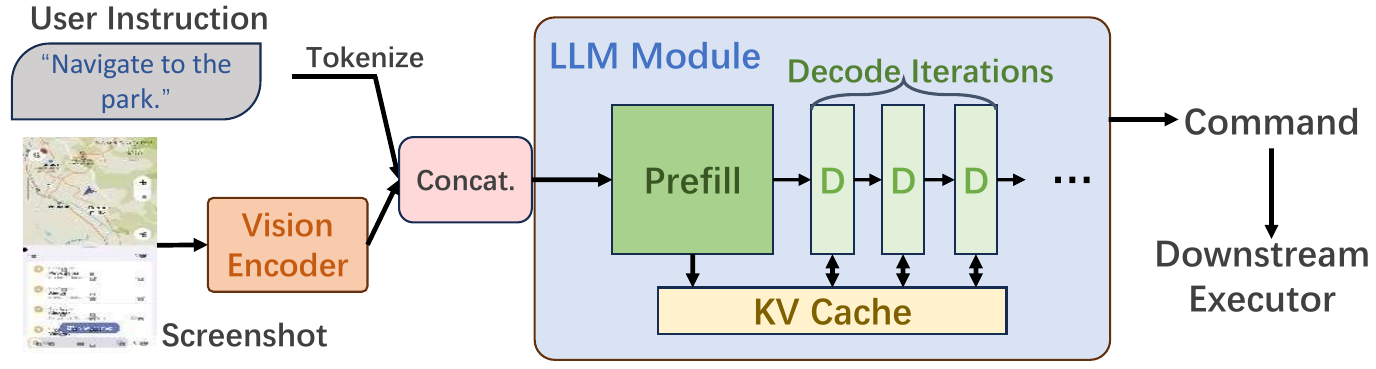}
    \caption{Framework of VLM-based GUI Agent.}
    \label{fig:vlm_agent}
\end{figure}

\begin{table}[t!]
\centering
\caption{Forward latency of different stages in VLM inference.}
\label{tab: stage_duration}
\resizebox{0.95\columnwidth}{!}{%
\begin{tabular}{@{}c|c|c|c@{}}
\toprule
Stage & Vision Encode & LLM Prefill & LLM Decode (Single Iter) \\ \midrule
Duration & 806.8 ms & 324.1 ms & 28.9 ms \\ \bottomrule
\end{tabular}%
}
\end{table}

A Vision-Language Model (VLM)-based agent is a multimodal large language model (LLM) designed to interpret graphical user interface (GUI) screens and generate executable actions based on natural language instructions~\cite{hong2024cogagent}. Its input typically includes a screenshot of the current GUI state and a user-issued instruction. The output is a structured action command, such as \texttt{<click button>}, \texttt{<scroll down>}, which an external controller can execute.

The architecture of a VLM-based agent, as shown in Figure~\ref{fig:vlm_agent}, consists of two main components: a vision module and an LLM module. The vision module uses a transformer-based encoder (typically CLIP-pretrained Vision Transformer (ViT)~\cite{radford2021learning_clip,dosovitskiy2020image_vit}) to transform the image into visual embeddings, which are adapted to the LLM’s input format by a lightweight adapter. Next, the LLM module concatenates the visual embeddings with the tokenized instruction and performs autoregressive generation to produce the textual responses (\ie sequence of text tokens). This process includes a prefill stage, where the multi-modal input is processed to build a key-value (KV) cache as context, and a decode stage, where tokens are generated one at a time sequentially based on the encoded context. The involvement of the KV cache improves computation efficiency by avoiding repetitive computation during sequential decode~\cite{kwon2023efficient_page_attn, yu2022orca}. 
As a reference, we report the forward latency of different stages in VLM inference in Table~\ref{tab: stage_duration}, testing using cogAgent~\cite{hong2024cogagent} on an NVIDIA RTX A6000 GPU. The results show that the forward latency of the vision stage is significantly longer than that of the prefill stage. Although a single decode iteration takes relatively little time, the decode stage still accounts for the majority of the total end-to-end latency (typically more than 50\%) due to the repeated execution over tens of token generation steps.

\textbf{Difference from General-Purpose VLMs:}  
To ensure the quality of decision making, agentic VLMs differ from general-purpose VLMs in the following aspect: Agentic VLMs typically accept higher-resolution images and employ larger model sizes, as shown in Table~\ref{tab:vlm_comparison}, to capture fine-grained visual details, perform precise pixel-level grounding, and generate reliable action commands for downstream execution. As a result, the vision encoder of agentic VLMs incurs significantly longer image processing time and places a substantially higher workload on limited GPU memory. 

\begin{table}[t!]
    \centering
    \caption{Comparisons of model parameters and input sizes of VLMs.}
    \label{tab:vlm_comparison}
    \resizebox{1.0\linewidth}{!}{
    \begin{tabular}{c|c|c|c|c}
    \toprule
        VLM & Type & LLM Size & VE Size & Image Size \\ \cmidrule{1-5}
        InternVL3-8B~\cite{chen2024internvl} & General & 7 Billions & 0.3 Billions &448$\times$448 \\ 
        LLAVA-Next-8B~\cite{li2024llava_next_interleave} & General & 8 Billions & 0.5 Billions & 672$\times$672 \\ 
        CogAgent-9B~\cite{hong2024cogagent} & Agentic & 9 Billions & 4 Billions & 1120$\times$1120 \\ \bottomrule
    \end{tabular}
    }
\end{table}

\subsection{Characterizing Agentic VLM Inference Workload}

The inference workload of a VLM-based GUI agent consists of three distinct stages: (1) vision encode, (2) LLM prefill, and (3) LLM decode. Each stage exhibits heterogeneous compute-memory characteristics that influence the hardware utilization and system scheduling strategy.

\paragraph{\textbf{VLM Computation Stages}}

\begin{table}[t!]
\centering
\caption{Resource utilization profiles of different stages, measured with Nsight Compute. The two values in each cell denote compute throughput (\%) and memory throughput (\%)\protect\footnotemark. }
\label{tab: resource_stage}
\resizebox{\columnwidth}{!}{%
\begin{tabular}{@{}c|ccccc@{}}
\toprule
\multirow{2}{*}{Stage} & \multicolumn{5}{c}{Kernel} \\ \cmidrule(l){2-6} 
 & \multicolumn{1}{c|}{\begin{tabular}[c]{@{}c@{}}Linear\\ (QKV)\end{tabular}} & \multicolumn{1}{c|}{\begin{tabular}[c]{@{}c@{}}Linear\\ (O)\end{tabular}} & \multicolumn{1}{c|}{\begin{tabular}[c]{@{}c@{}}Linear\\ (UG)\end{tabular}} & \multicolumn{1}{c|}{\begin{tabular}[c]{@{}c@{}}Linear\\ (D)\end{tabular}} & Attention \\ \midrule
Vision & \multicolumn{1}{c|}{83.5 / 57.0} & \multicolumn{1}{c|}{73.7 / 52.0} & \multicolumn{1}{c|}{87.7 / 59.2} & \multicolumn{1}{c|}{89.9 / 58.7} & 74.4 / 36.3 \\ \midrule
Prefill & \multicolumn{1}{c|}{74.5 / 58.5} & \multicolumn{1}{c|}{87.4 / 73.9} & \multicolumn{1}{c|}{92.3 / 58.9} & \multicolumn{1}{c|}{88.0 / 70.1} & 73.8 / 34.9 \\ \midrule
Decode & \multicolumn{1}{c|}{26.0 / 86.4} & \multicolumn{1}{c|}{26.6 / 88.2} & \multicolumn{1}{c|}{26.6 / 90.9} & \multicolumn{1}{c|}{27.8 / 92.3} & 17.0 / 53.1 \\ \bottomrule
\end{tabular}%
}
\end{table}
\footnotetext{\highlight{Both compute and memory throughput are aggregated metrics in Nsight Compute. In our setting, they correspond to the tensor pipeline active rate and the DRAM active rate, respectively.}}

Both the vision encode and LLM prefill stages exhibit compute-intensive behavior. The vision stage processes high-resolution GUI screenshots using a ViT model~\cite{dosovitskiy2020image_vit}, requiring heavy computation due to high image resolutions and model complexity. The LLM prefill stage corresponds to the initial forward pass of the language encoder with both visual and textual embeddings, involving dense matrix multiplications and attention over a long context, making the stage similarly compute-bound.
In contrast, the LLM decode stage is mostly memory-bound. It performs auto-regressive token generation, where the LLM predicts one token at a time using cached key-value (KV) pairs from the prefill stage. Although the input length during decode is much shorter, each iteration still requires loading the full LLM weights from GPU memory to on-chip compute units, \eg streaming multiprocessors (SMs) on NVIDIA GPUs, as well as accessing KV-cache, especially costly when serving multiple concurrent user requests. 

We report the resource utilization of four linear kernels and the attention kernel in the Transformer~\cite{vaswani2017attention_transformer} layers across the three stages using \texttt{nsight-compute}~\cite{nvidiaNVIDIANsight_nsight_compute}, as shown in Table~\ref{tab: resource_stage}. It can be observed that the decode stage exhibits significantly higher memory throughput compared to compute throughput, whereas the vision and prefill stages show the opposite trend, indicating highly heterogeneous resource demands across different stages.

\paragraph{\textbf{Agentic VLM Serving on a Single GPU}}  

We focus on deploying an agentic VLM serving system on an edge server with a single GPU, motivated by the high bandwidth cost of uploading screenshots and the need to preserve their privacy. Unlike cloud settings where VLM inference stages can be distributed across GPUs via high-bandwidth interconnects such as NVLink~\cite{li2019evaluating_nvlink, zhong2024distserve}, all stages in our setting run on a single GPU. 
Each request from a user device includes an instruction text (prompt) and a screenshot image. For multi-step tasks, the instruction may also contain historical interactions to maintain continuity across steps.

\highlight{Our serving system can be formulated as a multi-stage queuing system. The objective is to minimize the end-to-end latency of requests, which consists of the queuing delay and processing time at each stage. Both average and tail latencies affect the quality of service (QoS), with the latter primarily caused by prolonged queuing delays directly tied to system throughput. Since bounding the worst-case latency under unpredictable request arrivals is challenging, we focus on balancing per-request processing time and overall throughput, thereby achieving favorable average and tail latencies.}

\paragraph{\textbf{Scheduling Implications}} 
The heterogeneous characteristics of VLM inference introduce a fundamental trade-off between system throughput and per-request latency. Prioritizing the vision and prefill stages first, followed by batching decode requests, can significantly improve the overall request processing throughput by maximizing compute utilization and amortizing decode overhead, while reducing time-to-first-token latency (TTFT). However, this strategy may lead to increased or unstable time-between-token latency (TBT) in the decode stage (\ie interrupted token generation), which can degrade the user-perceived response smoothness. Therefore, achieving an optimal balance between overall throughput and per-request latency is a key challenge in scheduling agentic VLM workloads on a single GPU.

\subsection{Limitations and Opportunities of LLM Serving Systems} 

\subsubsection{\textbf{Limitations of Chunked Prefill for VLM Inference}}
Chunked prefill~\cite{agrawal2024taming_sarathi_chunkprefill} is a widely adopted technique in LLM serving systems aimed at reducing TBT and ensuring smooth generations. It works by splitting the user-provided prompt into smaller blocks, encoding them in multiple iterations, and combining the prefill requests and additional decode requests in a single batch, called hybrid batching. 
By doing so, it implicitly leverages the complementary resource demands of the prefill and decode stages to improve the overall GPU utilization. 
This strategy is effective in LLMs because prefill and decode stages share the same model weights, while Transformer architectures~\cite{vaswani2017attention_transformer} support batching of mixed operations, including token-wise computation in linear layers and optimized request-wise self-attention (\eg POD-Attention~\cite{kamath2025pod-attention}).

However, chunked prefill faces fundamental limitations when applied to agentic VLM inference. 
The vision encoder and LLM modules in VLMs correspond to separate model parameters, making it infeasible to combine the vision encode and LLM decode stages in a single batch. Moreover, the vision encode stage runs significantly longer than LLM prefill, further diminishing the benefits of chunked execution. 

\textbf{\underline{Insight 1}:} Existing LLM serving frameworks overlook the unique execution patterns of agentic VLMs and struggle to balance their overall throughput and per-request latency.

\subsubsection{\textbf{Opportunities in Cross-Stage Parallelization}}

\begin{figure}[t!]
    \centering
    \includegraphics[width=\linewidth]{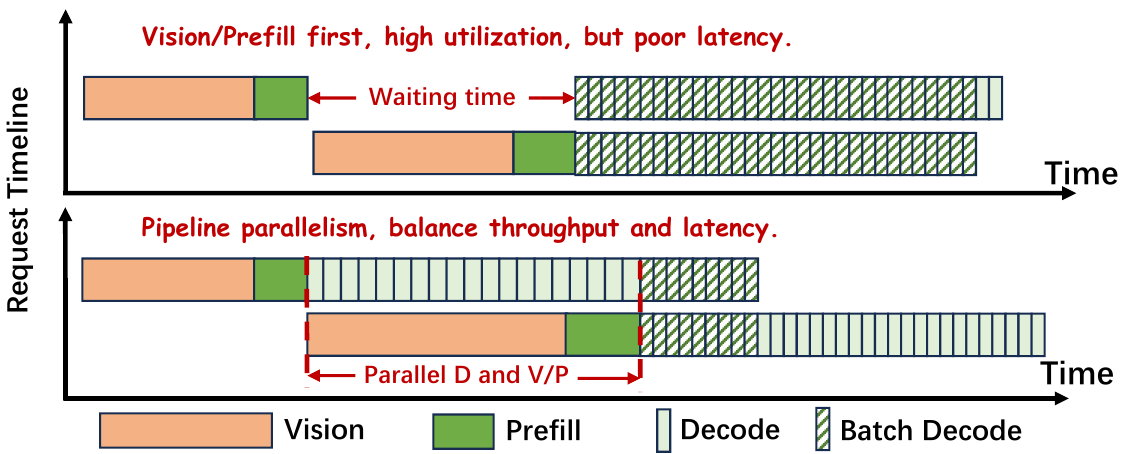}
    \caption{Timelines of two requests under different scheduling strategies. By leveraging pipeline parallelization between the decode and vision/prefill stages, the system achieves a better balance between GPU throughput and request-level latency.}
    \label{fig:pipeline_parallelization}
\end{figure}

Unlike operator-level parallelization techniques like chunked prefill, we argue that cross-stage pipeline parallelization is more effective for agentic VLM serving. 
First, it supports co-running across any inference stages without restrictions on their weight sharing. 
Second, it frees the scheduler from complex, manual request-level splitting and merging. 
Third, it better exploits the heterogeneous resource demands of different stages, leading to improved GPU utilization.

With properly designed pipeline parallelization across stages, it is possible to achieve a more favorable trade-off between throughput and latency. 
As illustrated in Figure~\ref{fig:pipeline_parallelization}, compared to traditional scheduling strategies that prioritize vision encode and LLM prefill stages, overlapping LLM decode with long-running vision or prefill stages eliminates part of the decode waiting time for the first request, thereby significantly reducing its TBT. 
Moreover, by intra-device parallelization and maintaining batching opportunities among decode requests, the impact on request processing throughput is reduced.

\textbf{\underline{Insight 2}:} Pipeline parallelization across VLM stages offers a superior balance between throughput and latency compared to rigid prefill-first or chunked prefill strategies.

\subsubsection{\textbf{Cross-Stage Parallelization and GPU Sharing}}
Efficient stage parallelization relies on fine-grained GPU sharing techniques. Recent works~\cite{strati2024orion, han2024kace} have explored GPU sharing via spatial multiplexing of low-level resources such as streaming multiprocessors (SMs), aiming to improve throughput by co-locating multiple kernels or tasks on a single GPU. However, most of these methods focus solely on maximizing resource utilization but fail to address the latency requirements of agentic VLM serving. Besides, integrating such low-level resource scheduling with the complex request-level coordination in serving systems poses additional challenges.

In agentic VLM serving, cross-stage GPU sharing plays two essential roles. First, it enables high-level request parallelism, helping to reduce overall end-to-end latency. Second, it promotes low-level GPU resource multiplexing by co-locating heterogeneous kernels from different stages. This co-location mitigates the throughput degradation caused by reduced batching opportunities in the decode stage under pipeline parallelism, as illustrated in Figure~\ref{fig:pipeline_parallelization}. Therefore, to achieve low latency across varying workloads, effective GPU sharing must account for both dynamic request fluctuations and the distinct compute and memory characteristics of each stage at the kernel level.

\textbf{\underline{Insight 3}:} Effectively combining GPU sharing techniques with VLM serving scheduling demands fine-grained, latency-aware resource coordination and remains a non-trivial problem.

\section{System Design}\label{sec:system_design}

\subsection{System Overview}
\begin{figure}[t!]
    \centering
    \includegraphics[width=0.95\linewidth]{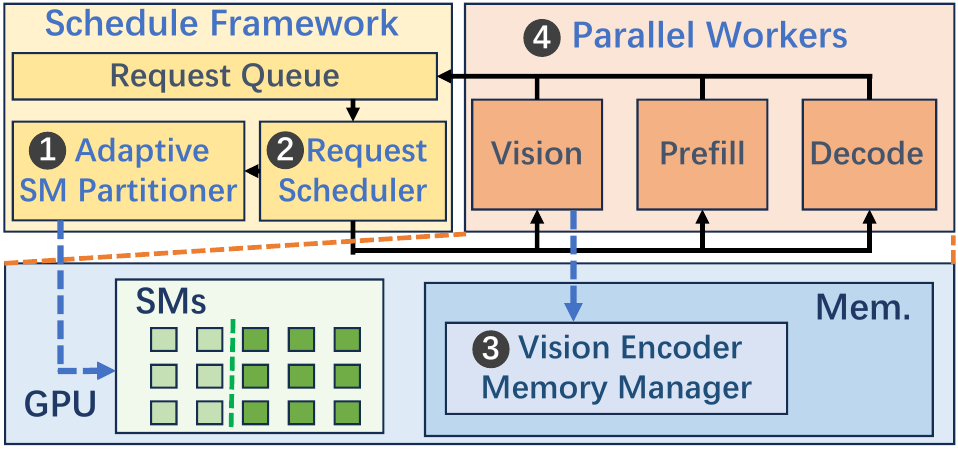}
    \caption{System Overview of \model.}
    \label{fig:framework}
\end{figure}

Figure~\ref{fig:framework} illustrates the overall architecture of \model. The system is designed to efficiently serve agentic VLM requests by optimizing GPU resource sharing, execution pipeline, and memory management. The key components and design principles are summarized as follows:

\begin{enumerate}[topsep=0pt,leftmargin=0.65cm,label=\protect\blackcircletext{\arabic*}]
    
    \item  \textbf{Adaptive GPU resource partitioner:} \model employs inter-SM (Streaming Multiprocessor) co-running to enable GPU spatial sharing between VLM stage workers. By co-locating compute-intensive and memory-bound kernels, and adaptively adjusting SM allocations based on workload, \model improves heterogeneous resource utilization and overall throughput.

   \item  \textbf{Runtime pipelined request scheduler:} A centralized request scheduler is responsible for tracking the states of active requests. It determines the request priorities and batching strategies among requests to balance per-request latency and overall throughput, particularly under varying runtime workloads. The scheduler runs in a separate control thread and communicates with each model worker via asynchronous message queues. 

   \item  \textbf{Vision encoder memory manager:} To address the high memory demands imposed by the large vision encoders in agentic VLMs, \model implements a layer-wise parameter swapping strategy. This technique efficiently loads vision encoder layers into GPU memory only when needed, thus leaving a significantly smaller memory footprint. 

    \item  \textbf{Parallel model workers:} The system consists of three dedicated model workers: a vision encode worker, an LLM prefill worker, and an LLM decode worker. These workers operate concurrently in separate threads. Each worker runs an event loop that continuously processes incoming requests dispatched by the central scheduler. Note that the LLM prefill and LLM decode workers share the same model parameters. 

\end{enumerate}

\subsection{Spatial GPU Sharing For Different Stages}

To enable parallel execution across VLM stages, we first explore GPU spatial sharing techniques that allow different kernels to corun efficiently. 
This section focuses on the underlying SM-level partitioning technique as a foundation for our adaptive scheduling strategy.

\subsubsection{\textbf{Infeasibility with multi-stream inference}} A straightforward approach to exploit spatial sharing across different inference stages is to use CUDA streams, which allow kernels from different streams to execute concurrently on the GPU. However, directly leveraging CUDA streams can lead to severe GPU resource contention and unpredictable latency fluctuations.

\begin{table}[t!]
\centering
\caption{Solo/Corun time and kernel-level statistics of two linear operators from the vision encoder and LLM decode.}
\label{tab:linear_corun}
\resizebox{\columnwidth}{!}{%
\begin{tabular}{@{}c|c|c|c|c|c|cc@{}}
\toprule
\multirow{2}{*}{Kernel} & \multirow{2}{*}{\begin{tabular}[c]{@{}c@{}}DRAM \\ Band.\\ (GB/s)\end{tabular}} & \multirow{2}{*}{\begin{tabular}[c]{@{}c@{}}SM \\ Throughput\\ (\%)\end{tabular}} & \multirow{2}{*}{\begin{tabular}[c]{@{}c@{}}Blocks\\ Per SM\end{tabular}} & \multirow{2}{*}{\begin{tabular}[c]{@{}c@{}}Total\\ Blocks\end{tabular}} & \multirow{2}{*}{\begin{tabular}[c]{@{}c@{}}Duration\\ (ms)\end{tabular}} & \multicolumn{2}{c}{\begin{tabular}[c]{@{}c@{}}Completion Time\\ (ms)\end{tabular}} \\ \cmidrule(l){7-8} 
 &  &  &  &  &  & \multicolumn{1}{c|}{Solo} & Corun \\ \midrule
Decode Linear & 639.78 & 33.58 & 1 & 448 & 0.35 & \multicolumn{1}{c|}{699.8} & 1241.8 \\ \midrule
Vision Linear & 273.14 & 87.63 & 1 & 3000 & 2.97 & \multicolumn{1}{c|}{588.3} & 680.6 \\ \bottomrule
\end{tabular}%
}
\end{table}

\begin{figure}[t!]
    \centering
    \includegraphics[width=0.98\linewidth]{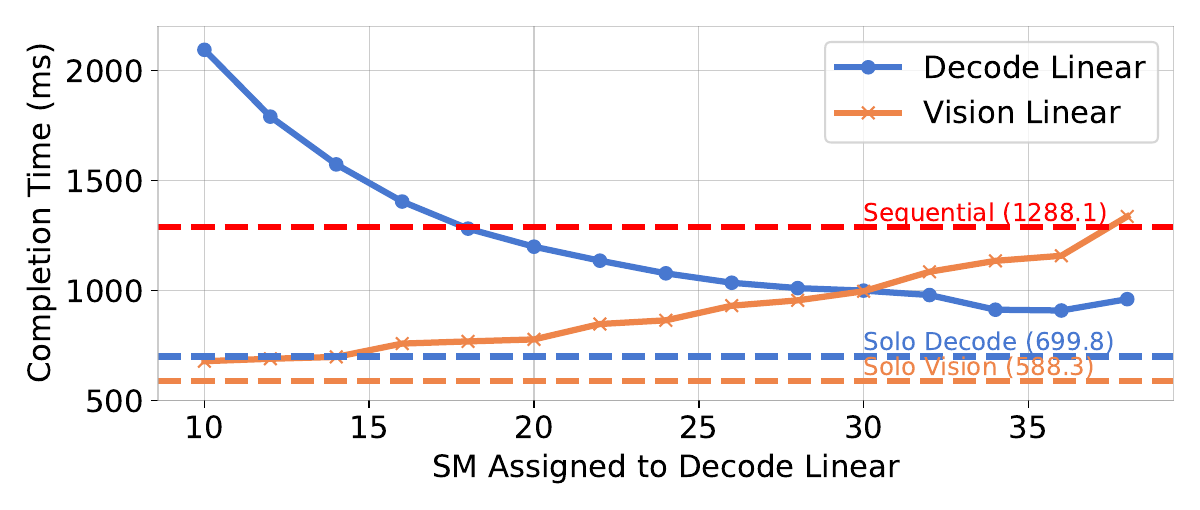}
    \caption{Kernel completion time varies with different SM allocations. Throughput improves over sequential execution when the decode linear kernel is assigned between 18 and 36 SMs.}

    \label{fig:linear_varying_sm}
\end{figure}

To demonstrate this, we select the two up-projection linear kernels—the largest linear kernels—from the vision encode and LLM decode, and execute them concurrently using CUDA streams on RTX A6000. The LLM decode linear kernel is repeated 2000 times, while the vision linear kernel is repeated 200 times. We also use \texttt{nvidia-compute} to profile the kernel-level statistics of these two operations. The results are summarized in Table~\ref{tab:linear_corun}. These two kernels are chosen because linear operations constitute the majority of execution time in both the vision encode and the LLM decode forward pass—approximately 75\% for LLM decode and 60\% for vision encode. The decode batch size is set to 3, as the output sequences of agent-oriented VLMs are typically short, which in turn limits the number of concurrent decode requests.

As shown in Table~\ref{tab:linear_corun}, the two linear operators exhibit different run-time characteristics: the decode linear kernel achieves high memory throughput, utilizing 83\% of the A6000's peak memory bandwidth (768~GB/s), but shows relatively low SM throughput—a metric that reflects compute intensity, primarily indicating tensor core utilization in this case. In contrast, the vision linear kernel is more compute-intensive. However, running them concurrently results in only marginal overall throughput improvement (1241.8~ms compared to 1288.1~ms for sequential execution), while the decode linear kernel experiences significant slowdowns.

The key reason for this performance limitation is severe \textbf{launch resource contention} between the two kernels. As shown in the table, both kernels launch significantly more thread blocks than the number of available SMs on the A6000 (84), which limits inter-SM parallelism. Additionally, each block consumes a substantial amount of per-SM resources (\eg registers and shared memory), allowing at most one block to be scheduled per SM. This further restricts intra-SM parallelism and reduces overall concurrency.

\subsubsection{\textbf{GPU spatial sharing via SM control}}To overcome this bottleneck, we adopt SM partitioning techniques to limit the number of SMs visible to each kernel, thereby ensuring that they can be launched concurrently. We use \texttt{libsmctrl}~\cite{libsmtrl_rtas} to control the SM allocation for different kernels, as it offers a more flexible and lightweight solution compared to other SM partitioning methods such as MIG~\cite{nvidia_mig} and Green Context~\cite{nvidia_green_context}. We vary the number of SMs assigned to the two linear kernels and use CUDA events to record their respective completion times, as shown in Figure~\ref{fig:linear_varying_sm}. 

By leveraging inter-SM parallelism, we observe throughput improvement: when the number of SMs allocated to the decode linear kernel is between 18 and 36, the completion time of the two co-running kernels is shorter than that of sequential execution. The main reason for this improvement is that memory-bound kernels like decode linear spend a significant portion of their execution time waiting for memory reads to complete. Since DRAM bandwidth and L2 cache—two critical resources for memory access—are shared among SMs, allocating too many SMs to memory-bound kernels leads to contention among thread blocks from the same kernel for memory access, causing frequent instruction stalls. 

Using \texttt{nvidia-compute}, we observe that over 40\% linear kernel execution cycles of LLM decode are stalled waiting for global memory access (as indicated by the Stall Long Scoreboard metric). As a result, its execution latency increases more slowly as the number of assigned SMs decreases, compared to compute-bound kernels like the vision linear. Therefore, co-running these two kernels can improve overall throughput. Besides, we also observe that in corun scenarios, the latency of both kernels is higher than their respective solo runs, even when sufficient SMs are allocated. This is due to contention for memory resources, such as DRAM bandwidth and L2 cache. Thus, inter-SM parallelism improves throughput at the cost of potential latency degradation.

\subsubsection{\textbf{Why not intra-SM parallelization for spatial sharing}} 
Many prior works~\cite{pai2013improving_elastic_kernel, zhao2021exploiting_intra_sm} leverage intra-SM parallelism to alleviate launch resource contention between GPU kernels. Inspired by these efforts, we apply two techniques to reduce the launch resource consumption of linear kernels: For vision linear, we use CUTLASS~\cite{Thakkar_CUTLASS_2023} to tune tile sizes and pipeline stages, significantly reducing register and shared memory usage with negligible performance degradation. For decode linear, which typically operate on small batch sizes, we reimplement them using CUDA without shared memory allocation and configure them to use CUDA cores instead of tensor cores, thereby avoiding compute unit contention with vision linear layers.
With these kernel-level optimizations, we observe throughput improvements even slightly exceeding that of inter-SM parallelism when co-running kernels. 

However, intra-SM parallelism depends heavily on the GPU’s internal scheduling behavior, which results in unstable kernel completion times due to the lack of control over kernel launch and collocation within the same SM. This instability persists even when combined with inter-SM techniques, making it hard to perform dynamic resource partitioning. Therefore, we argue that intra-SM parallelism is more suitable for throughput-oriented applications, but is less appropriate for our scenario, where a fine-grained balance between throughput and latency is critical. For this reason, we opt to only use the inter-SM parallelization strategy in this paper.

\subsection{Adaptive SM Partition} \label{subsec: adaptive_sm_partition}

\begin{figure}[t!]
    \centering
    \includegraphics[width=\linewidth]{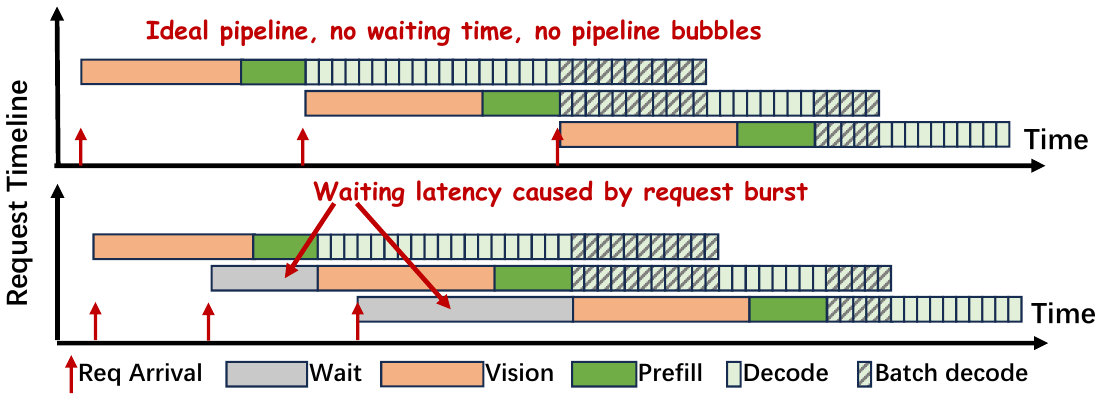}
    \caption{Waiting latency caused by request burst.}
    \label{fig:req_burst}
\end{figure}

Another important question is how to partition SMs among requests in different stages to achieve the best latency-throughput trade-off. We first consider this problem in an idealized scenario, \ie all requests are scheduled with perfect pipeline parallelism, without any waiting time in the vision encode/LLM prefill stages or pipeline bubbles, as illustrated in the upper example of Figure~\ref{fig:req_burst}. 

\subsubsection{\textbf{End-to-end request latency and SM partition}}
Assume the SM partitions for decode-vision and decode-prefill co-running are $P_{v}$ and $P_{p}$, respectively. Let the forward duration of LLM decode under these two co-running configurations be $t_d(P_{v})$ and $t_d(P_{p})$, and let $t_v(P_{v})$ and $t_p(P_p)$ denote the durations for the vision encode and LLM prefill stages when co-running with decode, respectively. \highlight{Assuming an token generation length of $L$, the expectation of the end-to-end latency $t_{e2e}$ in this ideal scenario is:
\begin{equation}
    \label{eq:e2e}
    \begin{aligned}
        \mathbb{E}[t_{e2e}] = \ &t_v(P_{v}) + t_p(P_{p})\  + \\ 
         & (prop_{v} \cdot t_d(P_{v}) + prop_p \cdot t_d(P_{p})) \cdot L,
    \end{aligned}
\end{equation} }
where $prop_v$ and $prop_p$ denote the proportion of co-running time with vision encode or LLM prefill during the entire decode iterations:
\begin{equation}
    \label{eq:prop}
    \begin{aligned}
        prop_v &= {t_v(P_v)}/{(t_v(P_v) + t_p(P_p))}, \\ 
        prop_p &= {t_p(P_p)}/{(t_v(P_v) + t_p(P_p))}.
    \end{aligned}
\end{equation}

After profiling the forward durations under different SM partitions and the average output length for token generation, we perform an enumeration search to determine the optimal SM partitions $P_v$ and $P_p$ that minimize the end-to-end latency in the ideal scenario, \ie
\begin{equation} 
    \label{eq: optimal_sm}
    \begin{aligned}
        &\min_{P_v, P_p} \ \mathbb{E}[t_{e2e}], \\  
        \text{s.t.} &\quad P_v, P_p \in \mathcal{P}_{\text{avai}},
    \end{aligned}
\end{equation}
where $\mathcal{P}_{\text{avai}}$ denotes the set of valid SM partitions\footnote{We only consider assigning SMs with continuous indices, as non-contiguous partitions offer no performance benefit.}. Assuming the total number of SMs is $N$, the time complexity of profiling is $O(N)$, and the complexity of searching for the optimal partition configuration is $O(N^2)$, which is lightweight given that modern GPUs typically have between tens and low hundreds of SMs.

\subsubsection{\textbf{Adaptive SM partition  strategy}}
\begin{figure*}[t!]
 \centering
 \subfigure[Latency-Throughput trade-off with Pareto front.\label{fig: pareto-a}]{\includegraphics[width=0.47\linewidth]{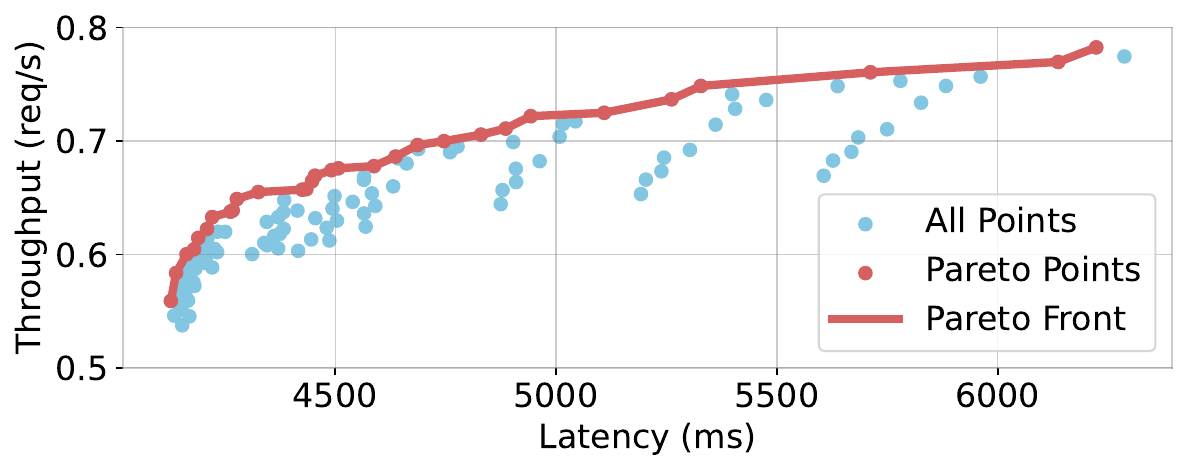}}
 \subfigure[\highlight{SM partition configurations under Pareto front.}\label{fig: pareto-b}]{\includegraphics[width=0.47\linewidth]{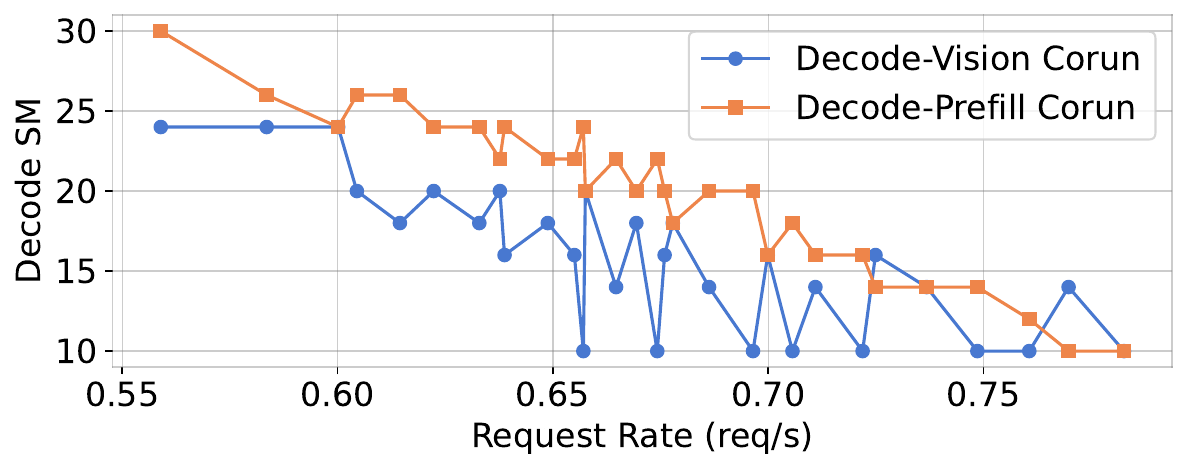}}
 \caption{Latency-Throughput trade-off under Pareto front.}
 \label{fig: pareto}  
\end{figure*}

However, static SM partitioning is inadequate for handling dynamic workloads, particularly during request surges. The optimal static configuration typically allocates relatively more SMs to the decode stage, though still fewer than those required to fully accelerate the vision encode or LLM prefill stages, to reduce iterative token generation latency. However, this leads to longer execution times for the vision encode and LLM prefill stages. As explained in the lower example of Figure~\ref{fig:req_burst}, during burst periods, the waiting time for requests in the vision encode and LLM prefill stages is significantly prolonged, as these stages exhibit no acceleration with request batching.

This motivates us to further analyze the throughput–latency trade-off introduced by SM partitioning. The end-to-end latency under a given SM partition configuration is given in Equation~(\ref{eq:e2e}), while the corresponding throughput $Thr$ can be approximated as:

\begin{equation}
    \label{eq:throughput}
    Thr = \frac{1}{t_v(P_v) + t_p(P_p)}.
\end{equation}
This is because the vision encode and LLM prefill are executed sequentially, whereas decode requests can be batched and executed in parallel with prefill and vision. Moreover, the latency of a decode iteration remains relatively stable as batch size increases. For instance, when the batch size increases from 1 to 10, the latency only rises slightly from 28.9\,ms to 30.6\,ms. In our serving scenarios, the decode batch size typically remains below 5 (reported in Section~\ref{sec:exper_overall}). Therefore, the ideal throughput is largely unaffected by the decode latency.

We report the latency and throughput under different SM partition configurations in Figure~\ref{fig: pareto-a}\footnote{\highlight{We use the average output length on the dataset, which are typically between 40–60 due to structured formatting in agentic VLM responses, justifying the use of the average.}}, and plot the Pareto frontier to identify the optimal trade-offs between latency and throughput. Furthermore, we illustrate the optimal SM partition configurations on the Pareto frontier corresponding to different throughput levels—\ie, request rates in serving scenarios—in Figure~\ref{fig: pareto-b}. It can be observed that as the request rate increases, the optimal strategy gradually allocates fewer SMs to the decode stage, following an approximately linear decreasing trend.

Since the real-time request rate is difficult to measure directly, we instead use the number of \textit{pending requests}, \ie those currently in the vision encode or LLM prefill stages, as an estimate of the current request rate and system load. Based on this, we design a dynamic strategy that adjusts the SM partitioning according to the number of pending requests: more SMs are allocated to the vision and prefill stages when the number of pending requests increases. The number of SMs assigned to the LLM decode stage is determined as follows:
\begin{equation}
    \label{eq:sm_assign}
    SM_{dec} = \max\left(SM_{min}, 
    SM_{op} - \alpha \cdot (N_{pend} - 1)\right),
\end{equation}
where $SM_{op}$ (calculated as equation (\ref{eq: optimal_sm})) denotes the default number of SMs allocated to the decode worker when only one request is active in the vision or prefill stage, and $N_{pend}$ is the number of requests currently in those stages. A minimum threshold $SM_{min}$ is enforced to bound the TBT and prevent excessive variability in token generation latency.

With this adaptive SM partitioning strategy, the scheduler can better balance overall throughput and per-request latency metrics. When the number of pending requests is high, \ie under heavy system load, reducing the resources allocated to decode and accelerating the processing of vision encode/LLM prefill stages provides two key benefits:
\begin{itemize}[topsep=0pt,leftmargin=0.45cm]
    \item It reduces the waiting time for requests in the vision encode and LLM prefill queues, preventing severe accumulation of pending requests.
    \item It allows earlier LLM decode requests to ``wait'' for later ones, thereby increasing opportunities for batching decode requests and improving overall system throughput, especially advantageous under high load conditions.
\end{itemize}

\subsection{Piepeline Request Scheduling}

\begin{algorithm}[t!]
\caption{Adaptive Request Scheduling Algorithm}
\label{alg:schedule}

\KwData{Request queue $Q$; Suspending decode request queue $Q_{d}$; Suspending vision request queue $Q_{v}$.}
\While{True}{
    req := get new request from $Q$\; 
    Update the number of requests in different stages\; 
    Adjust the SM partition according to eq~(\ref{eq:sm_assign})\;
    \lIf {req is finished} {send req to the output queue} 
    
    \lIf {req is prefill} {send req to the prefill worker}
    \If {req is decode} { 
                \lIf{there is decode running} {$Q_{d}$.append(Req)}
                \lElse {send Merge($Q_d$, req) to decode worker}
    } 
    \If {req is vision} {
        \lIf {no running vision and prefill requests} {\\ \ \ Send req to vision worker} 
        \lElse {$Q_{v}$.append(req)}
    }
}
\end{algorithm}

We now introduce our request pipeline scheduling algorithm with adaptive inter-SM spatial GPU sharing. The scheduler decides the priorities of different requests, the batch choices among requests, and the number of SMs assigned to model workers of different VLM stages to balance per-request latency and the overall system throughput. 

\textbf{Priority of Requests: } For requests within the same stage, we adopt a FIFO scheduling policy. Between stages, requests in the LLM decode stage are assigned the highest priority and are scheduled promptly. They can be executed in parallel with vision encode or LLM prefill requests to ensure consistent token generation and maximize GPU utilization. In contrast, vision encode and LLM prefill requests do not run in parallel, as both are compute-bound and would contend for GPU resources. We prioritize LLM prefill requests over vision encode requests because of shorter durations. Moreover, completing the LLM prefill stage enables the generation of new LLM decode requests, which can then be batched on the fly to improve overall throughput.

\textbf{Batch Decision:} We do not batch LLM prefill or vision encode requests, as both stages are highly compute-bound. Batching them does not improve throughput but instead increases the per-request latency. In contrast, we always batch LLM decode requests, whose batching significantly improves request throughput with minimal impact on per-request latency. Similar to existing LLM serving systems, we adopt an in-flight batching strategy~\cite{yu2022orca}. When a new LLM decode request arrives, if there is an ongoing LLM decode batch, the request waits until the current running batch finishes and then joins the next batch to be dispatched. This wait time is typically short, as LLM decode latency is generally low.

The request scheduling loop of \model is detailed in Algorithm~\ref{alg:schedule}. The request queue contains both newly arriving requests and those already being processed by model workers. The scheduler maintains two waiting queues: decode requests may be suspended to wait for forming an in-flight batch, and vision requests may be suspended to avoid blocking prefill requests. At the start of each scheduling iteration, the scheduler checks the status of active requests and adjusts the SM partition accordingly (Lines~2 to~4). For requests at different stages, the scheduler decides whether to dispatch them to model workers immediately or suspend them based on their priority and batching decisions (Lines~5 to~15).

We next describe the granularity of SM assignment calibration. Since the request scheduler and model workers run concurrently on separate threads, each SM re-assignment requires synchronization. Possible granularities include forward-pass-level (\ie adjusting SM allocation per forward pass) and kernel-level (\ie adjusting before each GPU kernel launch). While finer-grained control allows more precise partitioning, it also incurs higher synchronization overhead. As most kernels are short, frequent reassignments can cause significant CPU blocking and hinder asynchronous GPU execution. Therefore, we adopt a coarse granularity by adjusting SM partitions before each forward pass within the model worker.

\highlight{\textbf{Queueing Behavior:} Under Nova scheduling, an incoming request first waits in the vision worker’s queue. After vision encoding, it is immediately processed by the prefill worker without interruption from other vision-stage requests. The decode worker runs continuously, and a request only waits in its queue during the first iteration to join the in-flight batch—a typically negligible delay. Thus, the queueing delay is dominated by the vision worker’s queue, which can be modeled as a single-capacity queue, since requests in vision encoding and LLM prefill are not parallelized or batched. Assuming that request arrivals follow a Poisson process, the queueing delay can be modeled with an M/G/1 queue, where the service time $T$ consists of the vision encode and LLM prefill durations, \ie $T = t_v + t_p$. Under an arrival rate $\lambda$, the expected queueing delay $W_q$ is given by:
\begin{equation}
    \label{eq:waiting_latency}
    \mathbb{E}[W_q] = \frac{\lambda \, \mathbb{E}[T^2]}{2 \times (1 - \lambda \mathbb{E}[T])}.
\end{equation}
}

\subsection{Efficient Weight Offloading for Vision Encoder} 

As discussed in Section~\ref{sec:vlm_agent}, agentic VLMs have significantly larger vision encoder model sizes compared to general-purpose VLMs, causing higher memory pressure on GPU devices. Existing LLM serving systems typically preallocate memory for KV-cache storage~\cite{kwon2023efficient_page_attn}; thus, the enlarged model size reduces the available KV-cache capacity, potentially degrading both system throughput and serving quality. To address this challenge, we propose an efficient layer-wise weight offloading and swap-in strategy between GPU memory and CPU memory, leveraging compute and data transfer overlapping to minimize performance overhead.

\begin{figure}[t!]
    \centering
    \includegraphics[width=0.95\linewidth]{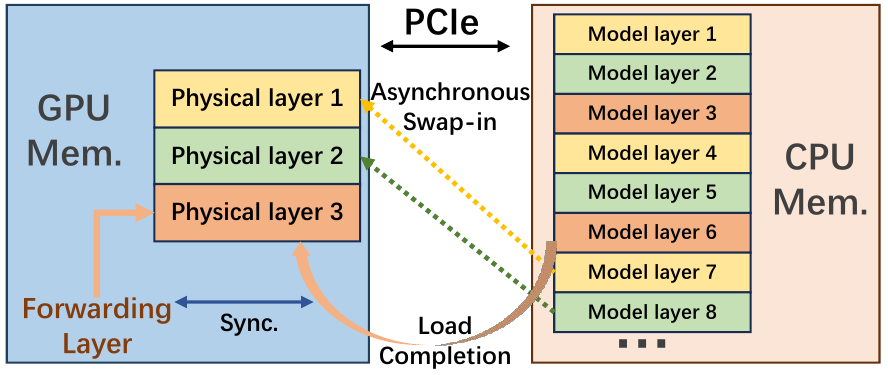}
    \caption{Weight swap-in between CPU and GPU memory. Physical layers and their corresponding model layers are marked with matching colors. The system is currently forwarding physical layer 3, which corresponds to model layer 6, while model layers 7 and 8 are being asynchronously swapped in.}
    \label{fig:weight_offload}
\end{figure}

Mainstream vision encoders typically adopt the ViT~\cite{dosovitskiy2020image_vit} architecture, comprising a stack of Transformer layers. Existing serving systems load all $L$ layers into GPU memory at initialization, leading to significant memory waste. In contrast, \model only allocates GPU memory for $K$ layers (called physical layers), where $2 \le K \ll L$, and initializes weights for only the first $K$ layers.
Each physical layer acts as a reusable buffer that holds the weights of multiple logical layers over time, as illustrated in Figure~\ref{fig:weight_offload}. During inference, once a model layer completes the forward pass, we asynchronously swap in the weights for the next logical model layer. Assume the current logical model layer ID stored in the physical layer is $cur\_layer$, the next logical layer to be loaded is:
\begin{equation}
nxt\_layer = (cur\_layer + K) \bmod L.
\end{equation}

Directly adopting weight offloading may incur significant memory transfer overhead. To reduce this, we apply three optimizations: (1) Forward computation and weight swap-in are placed in separate CUDA streams to enable compute–transfer overlap. (2) Pinned (page-locked) CPU memory is used for offloaded weights, allowing zero-copy DMA transfers between host and device. (3) Synchronization between the forward and weight-loading streams is coordinated using CUDA events. Specifically, each physical layer is associated with a CUDA event, and synchronization is performed before the layer is either executed or its weights are swapped in.

\textbf{Condition Analysis:} We now analyze the latency introduced by model weight loading and derive the memory bandwidth required for zero-overhead swap-in. Let $S$ be the total model size, $T$ be the single forward latency, and $B$ be the memory bandwidth. \highlight{During a forward pass, the first $K$ layers are preloaded into GPU memory. The initial preload occurs at system startup, while subsequent asynchronous swaps are overlapped with computation to hide their latency.}. When forwarding the $l$-th layer ($l > K$), at least $\frac{l - K}{L} \cdot S$ bytes must have been have been loaded into GPU memory. The available time window is $\frac{l - 2}{L} \cdot T$, since swap-in starts only after the first layer’s execution completes. To avoid stalling, the required memory bandwidth $B$ must satisfy:
\begin{equation}
    \label{eq:mem_band}
    \frac{l-2}{L}\cdot T \ge \frac{S}{B}\cdot\frac{l-K}{L}.
\end{equation}
\noindent This condition must hold for all $l \in [K+1, L]$, yielding that $B \ge \frac{S}{T} \cdot \frac{L - K}{L - 2} \approx \frac{S}{T}$ when $L \gg K$. The analysis for consecutive forward passes yields a similar result, indicating that the required bandwidth is independent of $K$. In practice, CogAgent’s vision encoder (8\,GB) takes 500--800\,ms per forward pass on an RTX A6000 or RTX 4090, requiring at most 16\,GB/s bandwidth---well below the typical peak bandwidth of PCIe 4.0$\times$16 (32\,GB/s), as commonly used for GPU interconnect. Therefore, the runtime weight swap-in introduces negligible latency overhead.


\textbf{Limitations and Potential Extensions:} Weight offloading is less effective for the LLM module due to the shared model weights between prefill and decode stages. Since each decode step is short, there is little opportunity to overlap weight loading with computation. However, prior work has proposed disaggregating prefill and decode~\cite{zhong2024distserve, patel2024splitwise}, allowing offloading to be applied on the prefill side. Additionally, on compute-rich but memory-constrained devices, increasing batch size can extend forward latency and help hide memory transfer overhead.

\section{Implementation}\label{sec:implementation}

We implemented the \model framework with  $\sim$2K lines of Python code, using PyTorch~\cite{paszke2019pytorch} as the backend library. The scheduler and the three parallel model workers run in separate threads and communicate via message queues. Each model worker performs forward passes in its own CUDA stream,  with all streams configured to the same priority level. SM partitioning is implemented by adding a Python interface to \texttt{libsmctrl}~\cite{libsmtrl_rtas}, assigning SMs to each stream with contiguous indices to each specific stage. We apply kernel fusion techniques for operators such as RoPE~\cite{su2024roformer_rope} and RMSNorm~\cite{zhang2019root_rmsnorm}, and use FlashInfer~\cite{ye2025flashinfer} for  attention kernels.

\section{Evaluation}\label{sec:experiment} 

\subsection{Experimental Setup}

\subsubsection{Hardware Platform}
We mainly conduct experiments on a server equipped with an AMD EPYC 7K62 48-core CPU, an NVIDIA RTX A6000 GPU, and 256\,GB of host memory. We also test \model on an RTX 4090 GPU. \highlight{The RTX A6000 features 84\,SMs and 48\,GB of VRAM, whereas the RTX 4090 offers 128\,SMs and 24\,GB of VRAM.}

\subsubsection{VLM Model}
We use cogAgent~\cite{hong2024cogagent}, a state-of-the-art agentic VLM. The model consists of a vision encoder, implemented as a 64-layer ViT~\cite{dosovitskiy2020image_vit} with 4B parameters, and an LLM module, which is a 40-layer Transformer model with 9B parameters. \highlight{The model precision is set to \texttt{bf16}, requiring approximately 9\,GB of GPU memory for vision encoding and 18\,GB for LLM prefill/decode. The KV-cache for a single request occupies about 70\,MB of GPU memory.}

\subsubsection{Dataset}
We adopt the Android Instruction Dataset from AndroidLab~\cite{xu2024androidlab}, which contains GUI interaction data collected from 138 tasks across nine Android applications running on virtual devices. Each sample includes a screenshot and a corresponding user instruction. The typical output length ranges from 30 to 80 tokens.

\subsubsection{Workloads}
Due to the lack of public agentic VLM serving traces, we simulate request arrivals using a Poisson distribution. By adjusting the parameter $\lambda$, we emulate various average request rates. Since the vision and prefill stages of each request take approximately 1.1\,s to complete, we limit the maximum average arrival rate to 0.8 requests per second to avoid severe system overload. Additionally, for experiments on the RTX 4090, we use private trace data collected from real-world serving scenarios to evaluate performance under practical workloads. To simulate varying load conditions, we scale the request arrival intervals of the trace to generate different request rates.

\subsubsection{Hyperparameter Setting}
 As introduced in Section~\ref{subsec: adaptive_sm_partition}, we determine $SM_{op}$ through enumeration based on profiling results under various SM partition configurations (reported in Section~\ref{subsec: experiment_sm_partition}). The optimal $SM_{op}$ values are 24 for decode–vision co-running and 30 for decode–prefill co-running. For both cases, we set $SM_{min}$ to 12, which ensures that the theoretical maximum TBT remains below 80\,ms. We set the $\alpha$ values for decode–vision and decode–prefill co-running to 4 and 6, respectively\footnote{\texttt{libsmctrl} only supports SM adjustments in units of 2, as each TPC (thread processing cluster) on the RTX A6000 consists of 2 SMs.}, meaning that $SM_{dec}$ will drop to $SM_{min}$ when the number of pending requests reaches 4. \highlight{The values of $\alpha$ are determined through offline profiling. We evaluated various $\alpha$ settings between 2 and 8 and observed stable performance across these configurations.}

\subsubsection{Metrics} We focus on two main categories of metrics.
\begin{itemize}[topsep=0pt,leftmargin=0.45cm]
    \item \textbf{Latency Metrics:} We primarily consider end-to-end (E2E) request latency, as it directly impacts the responsiveness perceived by downstream executors in agent scenarios. We report both the average and maximum E2E latencies observed in our experiments. Additionally, we report latency breakdown metrics, including time-to-first-token (TTFT) and time-between-tokens (TBT), as commonly adopted in standard LLM serving systems.
    \item \textbf{Throughput Metrics:} While prior works on LLM serving typically measure throughput using the rate of processed or generated tokens~\cite{zhu2024nanoflow}, this approach is less suitable for VLM agent serving, where the vision stage accounts for a significant portion of the total execution time but does not involve token generation. Therefore, we instead adopt request throughput (\ie number of processed requests per second) as our throughput metric.
\end{itemize}

\subsubsection{Baselines} 
\model is compared with three approaches:
\begin{itemize}[topsep=0pt,leftmargin=0.45cm]
    \item \textbf{PF-Limit:} Prefill-first scheduling is a widely adopted strategy in existing LLM serving systems such as vLLM~\cite{kwon2023efficient_page_attn} and SGLang~\cite{zheng2024sglang}, which prioritizes the vision encode and LLM prefill stages to maximize system throughput. Since naive prefill-first scheduling suffers from poor latency performance due to prolonged waiting times in the LLM decode queue, we introduce a threshold-based modification: LLM decode requests are scheduled when the number of waiting LLM decode requests exceeds a predefined threshold (set to 5 in our experiments).
    \item \textbf{Chunk~\cite{agrawal2024taming_sarathi_chunkprefill}:} Chunked prefill splits a LLM prefill request into several chunks and batches these chunks with LLM decode requests, thus reducing LLM decode latency. It also achieves better GPU utilization by locating compute-bound LLM prefill and memory-bound LLM decode requests in the same batch~\cite{vllmPerformanceTuning}. We evaluate various token budget settings and select 128 as the optimal value.
    \item \textbf{Multi-Stream:} Similar to \model, multi-stream scheduling allows kernels from different stages to execute concurrently on the GPU. However, it relies on CUDA's default multi-stream scheduling policy for GPU resource partitioning.
\end{itemize}

\subsection{Impact of SM Partition}
\label{subsec: experiment_sm_partition}

\begin{figure}[t!]
    \centering
    \includegraphics[width=0.98\linewidth]{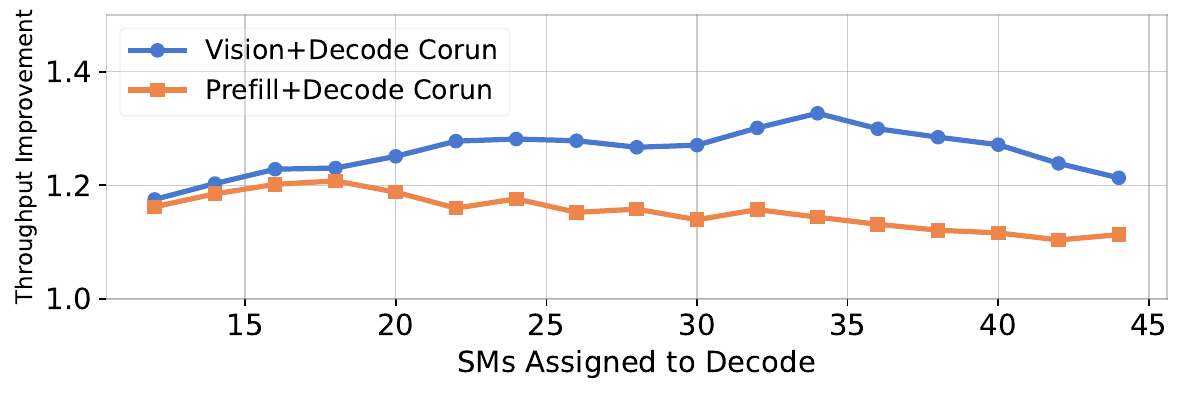}
    \caption{Corun throughput improvement under different SM allocations.}
    \label{fig:corun_throughput}
\end{figure}

\begin{figure}[t!]
    \centering
    \includegraphics[width=0.98\linewidth]{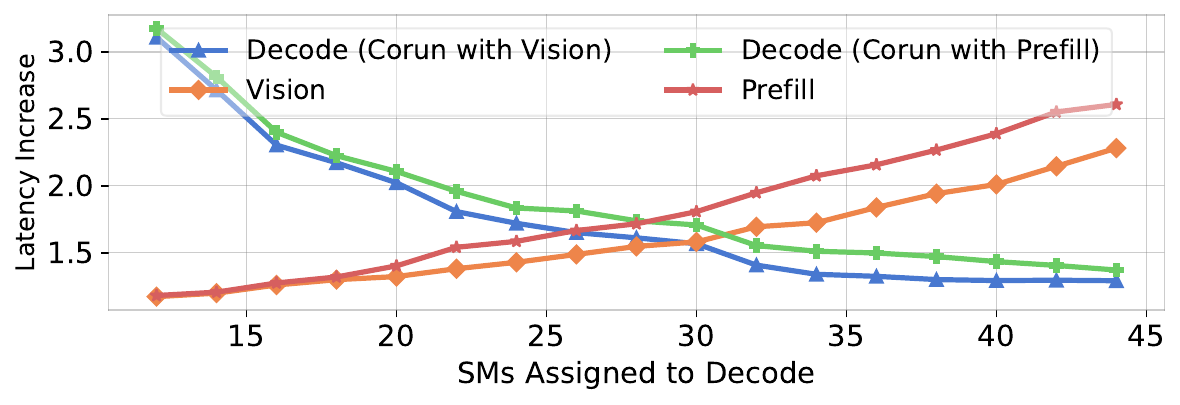}
    \caption{Corun latency increase under different SM allocations.}
    \label{fig:corun_latency}
\end{figure}

We first evaluate the performance of co-running requests from different stages under various SM partition configurations. We report both the overall throughput improvements and the corresponding single-pass latency increase introduced by co-running in Figure~\ref{fig:corun_throughput} and Figure~\ref{fig:corun_latency}. 
The main observations include: First, co-running LLM decode requests with either vision encode or LLM prefill requests improves overall throughput, with moderate variation across different SM allocations. We observe that its co-running with vision encode yields higher throughput improvement, because of the higher memory bandwidth contention between LLM prefill and LLM decode stages. Although the LLM prefill is shorter in duration than vision encode, it involves loading larger model weights, resulting in greater bandwidth usage.
Second, the latency increase for both decode–vision and decode–prefill co-running exhibits similar trends. However, the vision stage requires more SMs to maintain low latency, as it exhibits more compute-intensive properties.

\subsection{Overall Performance}
\label{sec:exper_overall}
\begin{figure}[t!]
    \centering
    \includegraphics[width=0.98\linewidth]{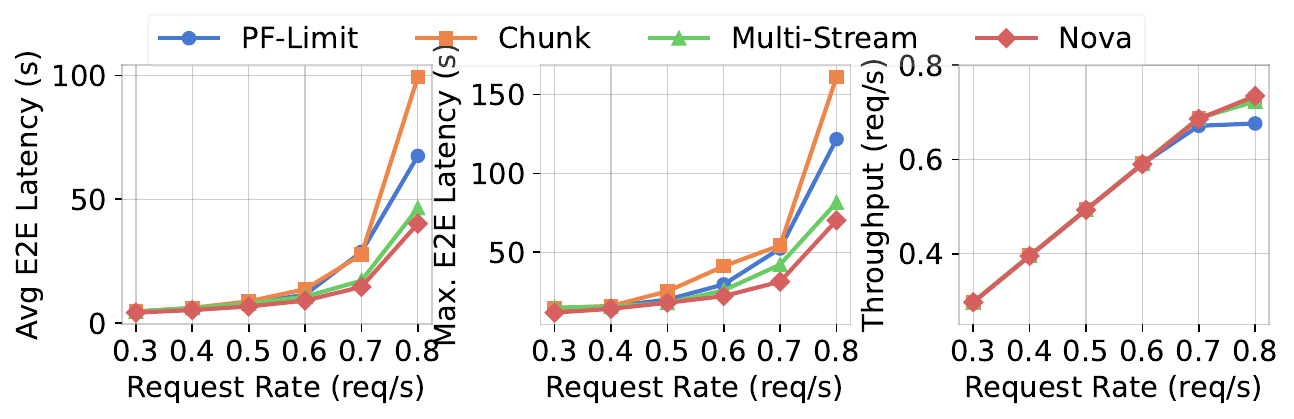}
    \caption{\highlight{E2E latency and throughput under different request rates.}}
    \label{fig:overall_effect_e2e_throughput}
\end{figure}

\begin{figure}[t!]
    \centering
    \includegraphics[width=0.98\linewidth]{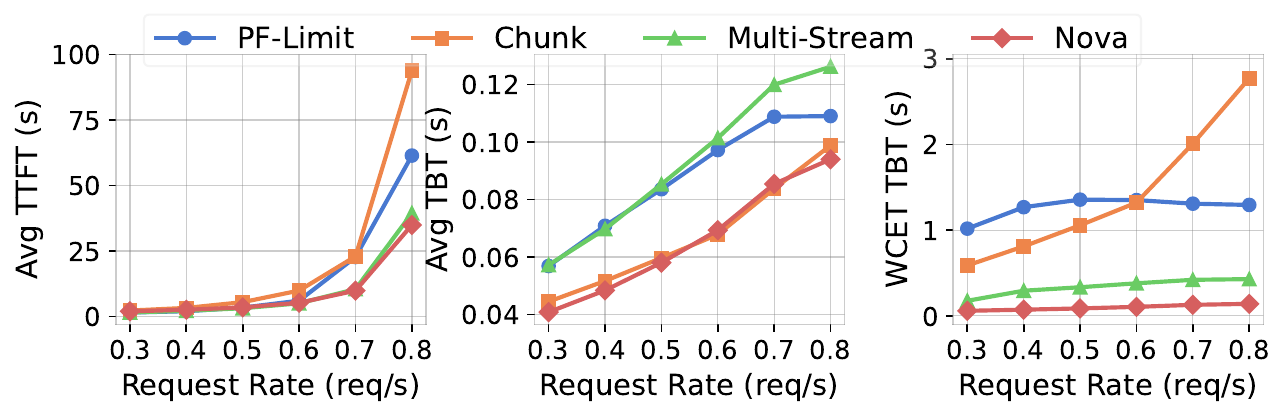}
    \caption{\highlight{Latency breakdown under different request rates.}}
    \label{fig:overall_latency_breakdown}
\end{figure}

Next, we compare the end-to-end performance under different request rates. \highlight{The request trace length is set to 500, spanning tens of minutes and capturing the stationary system behavior under sustainable request rates.} We report both end-to-end latency and throughput in Figure~\ref{fig:overall_effect_e2e_throughput}, and provide a latency breakdown in Figure~\ref{fig:overall_latency_breakdown}. 
\model demonstrates the best performance in both average and maximum end-to-end latencies, critical to downstream command executors. 
As shown in the latency breakdown, \model maintains a low TBT through steady token generation by parallelizing LLM decode requests with other request types. While chunked prefill enables parallelization between LLM prefill and decode stages, overlooking vision encode requests can still block LLM decode requests.
Multi-Stream relies on internal black-box GPU scheduling, tending to prioritize LLM prefill/vision encode over LLM decode as they launch significantly more thread blocks. 
The phenomenon results in severe blocking of decode execution, yielding low TTFT but high TBT.

\begin{figure}[t!]
    \centering
    \includegraphics[width=0.98\linewidth]{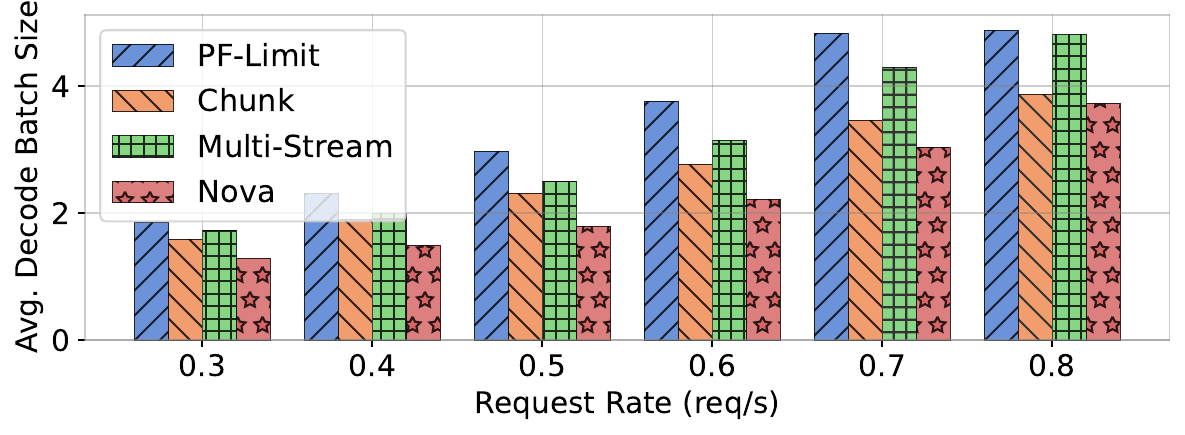}
    \caption{\highlight{Average decode batch size under different approaches.}}
    \label{fig:overall_decode_bs}
\end{figure}

To better understand the throughput gain brought by spatial gpu sharing, we report the average LLM decode batch sizes under different scheduling approaches in Figure~\ref{fig:overall_decode_bs}. Chunked prefill maintains throughput by creating more opportunities to batch LLM decode requests and batching LLM prefill chunks with LLM decode requests, thereby avoiding the SM under-utilization caused by solo LLM decode runs. 
\model exhibits the lowest average decode batch size due to its consistent token generation strategy, which keeps the number of active decode requests low. 
Nevertheless, it still achieves comparable throughput because of its integrated GPU spatial sharing strategy.

\highlight{We also report the queueing delays under Nova scheduling and compare them with the theoretical predictions from the M/G/1 model given in equation~(\ref{eq:waiting_latency}), as summarized in Table~\ref{tab: mg1_queue_latency}. The system utilization is defined as $\lambda T$. At a request rate of 0.8, the system is overloaded with utilization exceeding 1. It can be observed that the measured queueing delays closely match the theoretical predictions, indicating that Nova scheduling aligns well with the M/G/1 queue model. At a request rate of 0.7, the discrepancy is relatively larger, as the system is nearly saturated.}

\subsection{Impact of Weight Offloading}

Here we report the impact of weight offloading on the vision encoder. 
We evaluate on both the single forward pass and the end-to-end impact when integrated into request scheduling.
As summarized in Table~\ref{tab:layer_vision}, only 2 physical layers in GPU memory are sufficient to hide the latency of weight swapping, enabling over 90\% memory reduction for the vision encoder, with negligible overhead in both single-pass and end-to-end serving scenarios. Increasing the number of physical layers $K$ slightly raises the overhead, because of the increased synchronization complexity between model computing and weight swapping, as each physical layer requires a dedicated CUDA event for synchronization.

\begin{table}[t!]
\centering
\caption{\highlight{Measured queueing delays under Nova scheduling versus theoretical predictions from the M/G/1 model.}
}
\label{tab: mg1_queue_latency}
\resizebox{\columnwidth}{!}{%
\begin{tabular}{@{}c|cccccc@{}}
\toprule
Request Rate $\lambda$ & \highlight{0.3} & \highlight{0.4} & \highlight{0.5} & \highlight{0.6} & \highlight{0.7} & \highlight{0.8} \\ \midrule
Average Processing Time $T$  (s) & \highlight{1.39} & \highlight{1.44} & \highlight{1.43} & \highlight{1.42} & \highlight{1.38} & \highlight{1.35} \\ \midrule
System Utilization & \highlight{0.42} & \highlight{0.58} & \highlight{0.72} & \highlight{0.86} & \highlight{0.97} & \highlight{1.08} \\ \midrule
\begin{tabular}[c]{@{}c@{}}Average Measured\\  Queueing Latency (s)\end{tabular} & \highlight{0.60} & \highlight{1.14} & \highlight{2.06} & \highlight{3.82} & \highlight{8.49} & \highlight{33.57} \\ \midrule
\begin{tabular}[c]{@{}c@{}}Theoretical M/G/1 Queueing \\ Latency Prediction (s)\end{tabular} & \highlight{0.51} & \highlight{1.00} & \highlight{1.86} & \highlight{4.32} & \highlight{20.99} & \highlight{\textbackslash{}} \\ \bottomrule
\end{tabular}%
}
\end{table}

\begin{table}[t!]
\centering
\caption{\highlight{Weight offloading impact on compute and memory efficiency.} }
\label{tab:layer_vision}
\resizebox{0.95\columnwidth}{!}{%
\begin{tabular}{@{}cc|c|cccc@{}}
\toprule
\multicolumn{2}{c|}{\multirow{2}{*}{Model}} & \multirow{2}{*}{Raw} & \multicolumn{4}{c}{Vision Encode w/ Layer Offload} \\ \cmidrule(l){4-7} 
\multicolumn{2}{c|}{} &  & \multicolumn{1}{c|}{K=2} & \multicolumn{1}{c|}{K=3} & \multicolumn{1}{c|}{K=4} & K=5 \\ \midrule
\multicolumn{1}{c|}{\multirow{2}{*}{Single Forward}} & \begin{tabular}[c]{@{}c@{}}Forward Latency \\ (ms)\end{tabular} & 797.2 & \multicolumn{1}{c|}{798.6} & \multicolumn{1}{c|}{802.2} & \multicolumn{1}{c|}{804.8} & 805.8 \\ \cmidrule(l){2-7} 
\multicolumn{1}{c|}{} & \begin{tabular}[c]{@{}c@{}}GPU Mem.\\ Usage (MB)\end{tabular} & 8595.2 & \multicolumn{1}{c|}{692.1} & \multicolumn{1}{c|}{821.6} & \multicolumn{1}{c|}{951.1} & 1080.7 \\ \midrule
\multicolumn{1}{c|}{\multirow{2}{*}{Serving Scenario}} & \begin{tabular}[c]{@{}c@{}}Avg E2E \\ Latency (s)\end{tabular} & \highlight{11.96} & \multicolumn{1}{c|}{\highlight{11.85}} & \multicolumn{1}{c|}{\highlight{12.35}} & \multicolumn{1}{c|}{\highlight{12.21}} & \highlight{12.20} \\ \cmidrule(l){2-7} 
\multicolumn{1}{c|}{} & \begin{tabular}[c]{@{}c@{}}Maximum E2E \\ Latency (s)\end{tabular} & \highlight{21.94} & \multicolumn{1}{c|}{\highlight{21.83}} & \multicolumn{1}{c|}{\highlight{22.62}} & \multicolumn{1}{c|}{\highlight{22.39}} & \highlight{22.37} \\ \bottomrule
\end{tabular}%
}
\end{table}

\subsection{Contribution of Adaptive SM Partition} 

\begin{figure}[t!]
    \centering
    \includegraphics[width=0.98\linewidth]{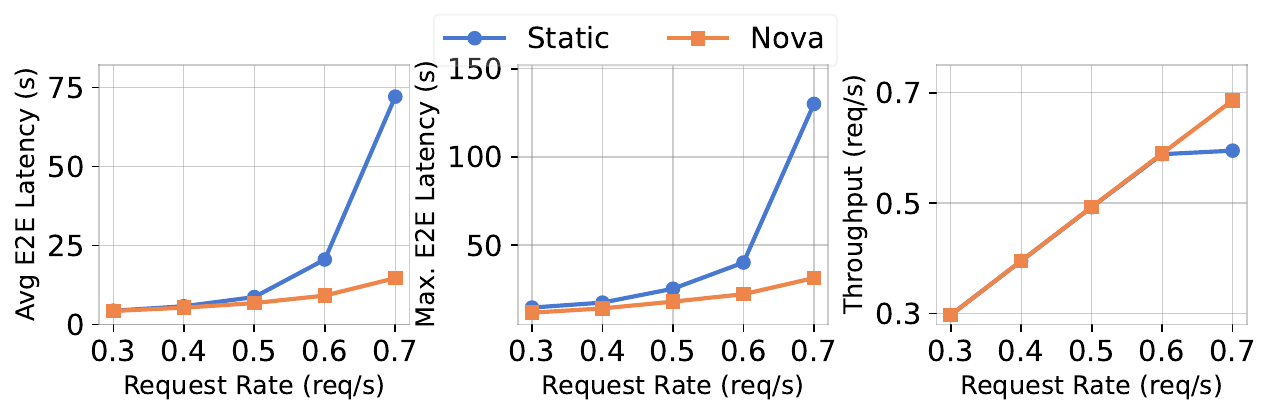}
    \caption{\highlight{E2E latency and throughput comparison between adaptive and static SM partition.}}
    \label{fig:adaptive_overall}
\end{figure}

We further analyze the adaptive SM partition strategy by comparing \model with static SM partition strategies, where fixed SMs are allocated to LLM decode requests when co-running with LLM prefill or vision encode stages. 
As shown in Figure~\ref{fig:adaptive_overall}, the proposed adaptive strategy consistently outperforms static counterparts across varying request rates, particularly when the request rate increases and request bursts become more frequent. Under high load, allocating too many SMs to LLM decode causes LLM prefill and vision encode requests to experience longer waiting times, further increasing end-to-end latency. 
In addition, static SM partitioning struggles to globally balance the system throughput between cross-stage co-running and LLM decode request batching. Over-allocating SMs to LLM decode reduces batching opportunities, as early LLM decode requests finish soon to wait for others.

\begin{figure}[t!]
    \centering
    \includegraphics[width=0.98\linewidth]{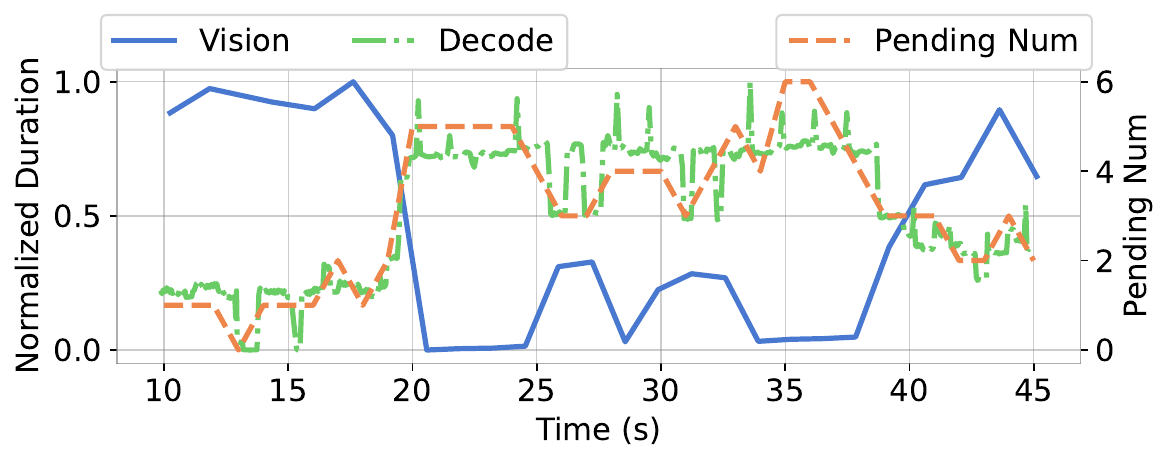}
    \caption{Practical adaptive SM partition example when requests burst.}
    \label{fig:adaptive_timeline}
\end{figure}

We also visualize the scheduling timeline of \model during a request burst in Figure~\ref{fig:adaptive_timeline}. The figure shows the latency changes of individual forward passes for LLM decode and vision encode requests, along with the fluctuation in the number of pending requests. Latency values are normalized between 0 and 1 for comparability. We find that when the number of pending requests increases, the scheduler dynamically allocates more SMs to vision encode requests to accelerate their processing. This adaptive behavior helps prevent the accumulation of pending requests and mitigates excessive waiting delays, ultimately improving overall system responsiveness.

\subsection{Scalability and Generalization Test} 
\begin{figure}[t!]
    \centering
    \includegraphics[width=0.98\linewidth]{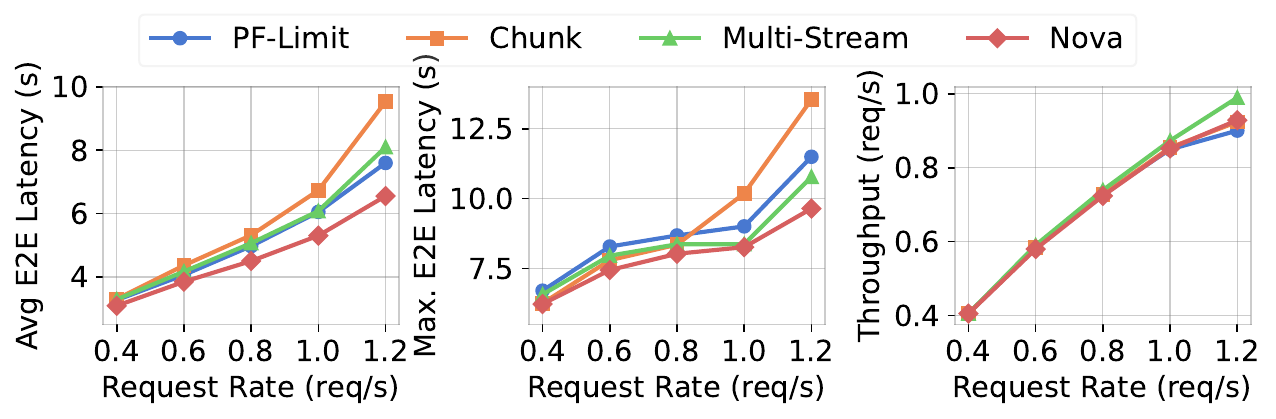}
    \caption{\highlight{Performance on RTX 4090 using real-world trace data.}}
    \label{fig:4090_experiment}
\end{figure}

To evaluate the generalization of \model into GPU models with smaller memories and real-world request distributions, we conduct additional experiments on an RTX 4090 using real-world traces from a production VLM agent, whose results are given in Figure~\ref{fig:4090_experiment}. 
Note CogAgent has 13\;B model parameters, and the model alone consumes approximately 27\;GB of GPU memory in BF16 precision, exceeding the 24\;GB memory capacity of RTX 4090. 
Therefore, we adopt weight offloading for the vision encoder in all approaches. The results show a similar trend to those on RTX A6000: \model consistently outperforms other baselines in terms of end-to-end latency while maintaining comparable throughput, demonstrating its scalability and cross-platform compatibility.

\subsection{Overhead Analysis}

\begin{figure}[t!]
    \centering
    \includegraphics[width=0.98\linewidth]{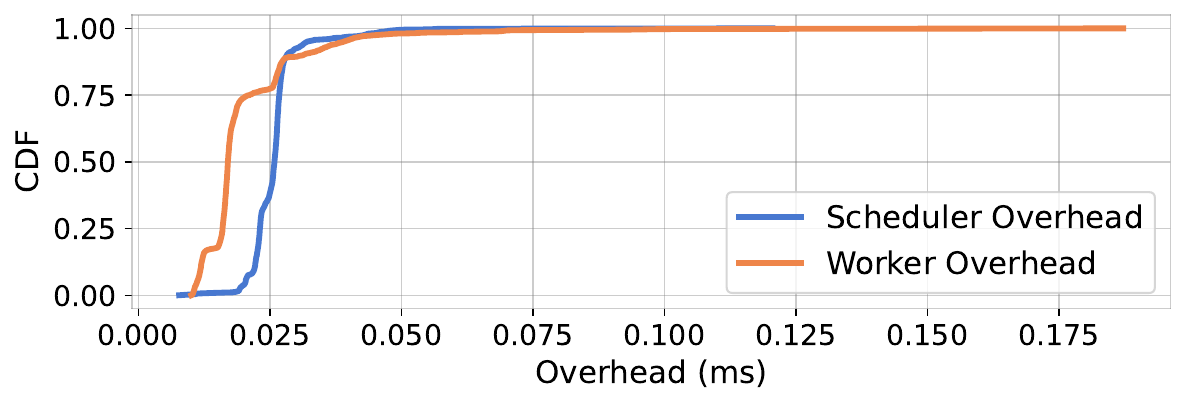}
    \caption{Overhead CDF of scheduler and model worker.}
    \label{fig:overhead}
\end{figure}

We finally report the scheduling overhead of \model, which mainly comes from the synchronization between the scheduler and model workers for SM partition adjustments. As shown in Figure~\ref{fig:overhead}, we present the cumulative distribution function (CDF) of overhead. The overhead remains lightweight compared to the overall model inference latency. Specifically, the model worker performs synchronization and performs the SM re-partition only once before each model forward pass. Moreover, since the scheduler runs in a separate thread and the model worker continues inference concurrently during request scheduling, the scheduling overhead is further minimized.

\section{\highlight{Discussion}}\label{sec: discussion}

\highlight{


\textbf{Extensions to Additional VLMs and Edge Devices:} Typical VLMs and other multi-modal LLMs use separate modality encoders with LLM modules~\cite{multi-modal-llm, shu2023audio_llm}. Our hardware-level spatial sharing and pipelining approach is broadly applicable: Prefill-decode co-running remains effective, while vision-decode co-running benefits depend on model size and image resolution, which affect hyperparameter tuning.
When deploying large-scale VLMs on resource-constrained devices, the efficiency of weight offloading is primarily determined by the ratio between device compute capability and main memory bandwidth. Our experiments indicate that PCIe~3.0 bandwidth is sufficient to hide weight transfer latency. Nevertheless, further improvements in model quantization~\cite{frantar2022gptq, lin2024awq, dettmers2022gpt3_int8} are needed, since quantizing the vision encoder can lead to significant accuracy degradation. Complementary strategies, such as self-speculative decoding~\cite{zhou2024survey_llm_inference, zhang2023draft_self_spec, elhoushi2024layerskip_self_spec}, may also be employed to improve LLM inference efficiency.

\textbf{Exploiting Frame Similarity:} One promising direction to improve efficiency is exploiting frame similarity across requests, especially when \model is employed in video processing with high temporal redundancy. Currently, \model does not use similarity-based batching or region skipping, as cross-attention over all image patches or tokens makes reusing intermediate features potentially harmful to accuracy.
However, there are two potential ways to exploit. 
First, the vision encoder can be accelerated by skipping redundant regions with high similarity, for example by merging or pruning tokens that are highly similar~\cite{song2024cmc_vit, lee2023multi_vision_similarity, black2022visualizing}. 
Second, the KV-cache of image tokens corresponding to similar regions can be reused in the LLM, while carefully retaining important tokens for recomputation to preserve accuracy~\cite{yao2025cacheblend, yang2025kvlink}.
}

\section{Related Works}\label{sec:related}

\textbf{Efficient LLM Serving Systems.} Prior work on LLM serving falls broadly into three categories. 
First, \textit{memory efficiency}, particularly targeting at alleviating KV-cache bottlenecks~\cite{kwon2023efficient_page_attn, rtss_demand_layer, zheng2024sglang, rehg2024kv_compress, lee2024infinigen, lin2024infinite_llm, li2024flexnn, kang2024rt_swap, giannessi2024rt_malloc, prabhu2025vattention}. For example, PagedAttention~\cite{kwon2023efficient_page_attn} reduces fragmentation via block-level cache management, while SGLang~\cite{zheng2024sglang} improves cache reuse with RadixCache.
Second, \textit{GPU utilization and throughput}, targeting better resource efficiency~\cite{agrawal2024taming_sarathi_chunkprefill, kamath2025pod-attention, zhong2024distserve, yu2022orca, patel2024splitwise, zhu2024nanoflow, rtas_gpu_manage, chen2025pre}. DistServe~\cite{zhong2024distserve} disaggregates prefill and decode across devices, and NanoFlow~\cite{zhu2024nanoflow} overlaps heterogeneous kernels via fine-grained mini-batching.
Third, \textit{quality of service}, which aims to meet latency metrics such as TTFT and TBT~\cite{cheng2024towards_slo_optimized, lin2024planck_slo, li2023rt_lm, wang2023progressive}. 
Despite these efforts, most approaches overlook the distinct execution patterns of multi-modal LLMs, especially agentic VLMs. Specifically, they do not account for the heterogeneous compute and memory demands, nor the stage-level modularity, present in VLM agent workloads. To the best of our knowledge, no dominant serving framework has yet been proposed to systematically address the challenges of efficient and low-latency agentic VLM inference.

\textbf{GPU Resource Sharing.}  
GPU sharing techniques aim to improve utilization and support concurrent execution. For DNN inference, both temporal and spatial sharing have been explored to reduce idleness and enable parallelism~\cite{strati2024orion, han2024kace, wu2023transparent_tgs, xiao2020antman, yu2020fine_salus_share, han2022microsecond_reef, xu2024flex}, typically by leveraging heterogeneous resource demands for co-scheduling.
Low-level GPU partitioning has also been studied to support kernel co-execution~\cite{chow2023krisp, zhao2021exploiting_intra_sm, libsmtrl_rtas, pai2013improving_elastic_kernel, gupta2012study_persistent_thread, nvidia_mig, nvidia_mps, nvidia_green_context, jain2019fractional_gpu}, and can be categorized into inter-SM and intra-SM approaches. Intra-SM techniques, such as elastic kernels~\cite{pai2013improving_elastic_kernel, zhao2021exploiting_intra_sm}, provide fine-grained sharing but often suffer from unstable performance due to opaque scheduling.
Moreover, intra-SM partitioning tends to prioritize throughput improvements without providing sufficient flexibility for fine-grained, latency-aware scheduling, making it less suitable for dynamic, multi-stage VLM serving workloads. In contrast, inter-SM partitioning offers more stable performance and better compatibility with adaptive scheduling policies.
Therefore, we adopt inter-SM sharing techniques to support adaptive, latency-aware resource allocation in VLM agent serving.

\section{Conclusion}\label{sec:conclusion}

We presented \model, a real-time scheduling framework tailored for multi-stage agentic vision-language model serving on a single GPU.
Nova integrates adaptive cross-stage pipeline parallelization with fine-grained SM partitioning to fully exploit GPU resources, and employs a lightweight weight offloading strategy to alleviate memory constraints by high-resolution vision encoders. 
Through dynamic scheduling based on a latency–throughput Pareto frontier, Nova ensures consistent responsiveness under diverse and bursty workloads. 
Extensive experiments on both synthetic and real-world agent tasks demonstrate Nova’s superiority in terms of end-to-end latency, throughput, and GPU memory efficiency.

\section*{Acknowledgement}
This work was sponsored in part by the National Key R\&D Program of China (No. 2022ZD0119100), in part by China NSF grant No. 62472278, 62025204, 62432007, 62441236, 62332014, and 62332013, in part by the Taishan Industrial Experts Program, in part by Alibaba Group through Alibaba Innovation Research Program, and in part by Tencent Rhino Bird Key Research Project. 
This work was partially supported by SJTU Kunpeng \& Ascend Center of Excellence.
The opinions, findings, conclusions, and recommendations in this paper are those of the authors and do not necessarily reflect the views of the funding agencies or the government.

\balance

\newpage
\bibliographystyle{IEEEtran}
\bibliography{references}
\vspace{12pt}

\end{document}